\documentclass{article}
\usepackage{PRIMEarxiv}
\usepackage{microtype}

\usepackage[utf8]{inputenc} 
\usepackage[T1]{fontenc}    
\usepackage{hyperref}       
\usepackage{url}            
\usepackage{booktabs}       
\usepackage{amsfonts}       
\usepackage{nicefrac}       
\usepackage{lipsum}
\usepackage{fancyhdr}       
\usepackage{graphicx}       
\graphicspath{{media/}}     
\usepackage{amsmath}
\usepackage{setspace}
\title{The Exploratory Study on the Relationship Between the Failure of Distance Metrics in High-Dimensional Space and Emergent Phenomena
}
\author{
    \begin{tabular}{ccc} 
       \large\textbf{HongZheng Liu} & \large\textbf{YiNuo Tian} & \large\textbf{Zhiyue Wu} \\ 
        \small Independent & \small Independent & \small Independent \\ 
        \footnotesize China & \footnotesize China & \footnotesize China \\ 
        \footnotesize {weiyouyeyu}@foxmail.com & \footnotesize {tianyinuo}282@gmail.com & \footnotesize {Zhiyuewu70}@gmail.com \\ 
    \end{tabular}
}
\usepackage[numbers,square]{natbib}
\begin{document}
\maketitle
\begin{abstract}
This paper presents a unified framework, integrating information theory and statistical mechanics, to connect metric failure in high-dimensional data with emergence in complex systems. We propose the "Information Dilution Theorem," demonstrating that as dimensionality ($d$) increases, the mutual information efficiency between geometric metrics (e.g., Euclidean distance) and system states decays approximately as $O(1/d)$. This decay arises from the mismatch between linearly growing system entropy and sublinearly growing metric entropy, explaining the mechanism behind distance concentration. Building on this, we introduce information structural complexity ($C(S)$) based on the mutual information matrix spectrum and interaction encoding capacity ($C'$) derived from information bottleneck theory. The "Emergence Critical Theorem" states that when $C(S)$ exceeds $C'$, new global features inevitably emerge, satisfying a predefined mutual information threshold. This provides an operational criterion for self-organization and phase transitions. We discuss potential applications in physics, biology, and deep learning, suggesting potential directions like MI-based manifold learning (UMAP+) and offering a quantitative foundation for analyzing emergence across disciplines.
\end{abstract}
\keywords{High-dimensional data \and metric failure \and complex systems \and emergence \and information theory  \and  statistical mechanics  }

\section{Introduction}
Over the past few decades, high-dimensional data analysis and complex systems research have garnered widespread attention across various disciplines, including statistical physics, machine learning, and systems biology. 
With the advent of the era of big data, the widespread application of high-dimensional datasets (such as gene expression data, image features, and text embeddings) has exposed severe limitations of traditional geometric metrics in high-dimensional spaces, commonly referred to as the "breakdown of distance metrics" or the concentration of distances problem \cite{ref24,ref35}. 
Meanwhile, in complex systems, whether in physical phase transitions, the formation of hierarchical features in neural networks, or the evolution of gene regulatory networks and public opinion dynamics, there is a common occurrence of spontaneously emerging global features under specific critical conditions—phenomena collectively referred to as emergence \cite{ref3}. 
Although these two categories of problems have been extensively studied in their respective domains, a unified theoretical framework explaining their intrinsic connection remains absent. 
The problem of distance concentration in high-dimensional data analysis was first identified by Beyer et al. \cite{ref17} and later substantiated both theoretically and empirically by Aggarwal and others. 
As dimensionality increases, traditional linear metrics such as Euclidean distance and cosine similarity tend to homogenize, making distinctions between data points increasingly difficult, thereby significantly impairing the effectiveness of methods like nearest-neighbor search, clustering, and manifold learning \cite{ref17,ref24}. 
At the same time, research in complex systems has shown that when a system approaches a critical state, coupling between local variables intensifies dramatically, leading to the spontaneous formation of novel macroscopic ordered structures or functional modules—a phenomenon broadly referred to as emergence \cite{ref3,ref4}. 
While significant progress has been made in explaining emergence through theories of phase transitions, network coupling models, and statistical physics, these approaches often emphasize aspects such as energy, interactions, or network topology, without providing a systematic explanation of emergence from the perspective of information transmission and encoding \cite{ref34,ref35}. 
Despite the rich body of work in each respective field, a unified explanation of these phenomena within the framework of information theory remains lacking \cite{ref3}. 
This gap forms the starting point of the present study: emergence is a profound and captivating phenomenon. 
Much like Brownian motion in the early twentieth century \cite{ref41} - an initially marginalized experimental observation that ultimately gave rise to the birth of statistical physics \cite{ref4}—today's study of complex systems stands at a similar crossroads. 
From the breakdown of distance metrics in high-dimensional space \cite{ref24,ref55,ref56}, the microscopic origin of entropy in black hole thermodynamics\cite{ref49,ref50}, and the collective motion of active matter \cite{ref43,ref51,ref52}, to the spontaneous emergence of hierarchical features in deep neural networks \cite{ref12,ref41,ref42,ref53,ref54}, these cross-disciplinary modern Brownian motion phenomena are widely observed but lack a unified theoretical explanation. 
They collectively suggest that when the structural complexity of a system's information exceeds the encoding capacity of interacting agents, the emergence of new order becomes an inevitable result of this fundamental contradiction. 
This paper proposes the construction of a unified information-theoretic–statistical-mechanical framework, aiming to quantitatively analyze the essence of metric failure in high-dimensional data from the perspective of mutual information, while also revealing the mathematical underpinnings of emergence in complex systems. 
The main contributions of this study can be summarized as follows: 
\begin{enumerate} 
\item \textbf{Information Dilution Theorem}: This paper defines a metric for mutual information efficiency $\eta(D) = \frac{I(D;S)}{H(S)}$, where $D$ denotes the Euclidean distance or its weighted variants, and $S$ represents the system state composed of multiple random variables. Through rigorous theoretical derivation, we prove that in high-dimensional systems composed of independent or weakly correlated variables, the total system entropy $H(S)$ grows linearly with the dimension $d$, while the entropy of metric $D$ increases only at a rate of $O(1)$. This leads to a mutual information efficiency decay of $O(1/d)$. This theorem reveals the intrinsic mechanism behind the "concentration of distances" phenomenon from an information-theoretic perspective and provides a theoretical foundation for metric selection in high-dimensional data processing. 
\item  \textbf{Emergence Critical Theorem}: Based on the spectral decomposition of the system's mutual information matrix, we introduce the concept of interaction encoding capacity $C$ to characterize the maximum mutual information that can be preserved by low-dimensional representations. We also define the information structural complexity $\mathcal{C}(S)$, which describes the number of independent information interaction modes within the system. When the information structural complexity of a system exceeds its encoding capacity, traditional low-dimensional representations fail to capture the full scope of system information, inevitably giving rise to new global features $f_{\text{new}}$, thus leading to the emergence phenomenon. Using information bottleneck theory and Lagrangian optimization, we quantitatively characterize this process and provide a sufficient condition for emergence in the form $\frac{I(f_{\text{new}};S)}{H(S)} > \theta$, where $\theta$ is a pre-specified significance threshold. This theoretical result offers a novel and operational mathematical criterion for explaining emergence in complex systems. 
\end{enumerate}
To achieve the above objectives, this study primarily adopts a methodology combining theoretical derivation and mathematical proof. First, starting from fundamental concepts in information theory, we rigorously derive the decay law of mutual information efficiency in high-dimensional systems and validate the theoretical result through detailed derivations for systems of i.i.d. Gaussian variables (see Appendix \ref{app:D}) \cite{ref24}. 
Subsequently, with the aid of information bottleneck theory and spectral decomposition techniques, we construct the theoretical framework of the Emergence Critical Theorem and extend the proof to weakly correlated and nonlinear systems in Appendix \ref{app:E}. 
In addition, we explore the mechanism of information emergence in artificial neural networks (see Appendix \ref{app:C}) and propose novel ideas such as “UMAP+” \cite{ref29}. 
The structure of the paper is arranged as follows: Section \ref{section3} introduces the basic principles of information theory and the background of metric failure in high-dimensional spaces, where we derive and prove the Information Dilution Theorem and the Emergence Critical Theorem, and interpret their physical implications. 
Section \ref{section4} discusses the applicability of the theoretical framework to physical phase transitions, biological network differentiation, and deep neural networks. It proposes improved manifold learning algorithms based on mutual information and information-topological analysis methods, and outlines interdisciplinary paths for empirical validation. 
Section \ref{section5} summarizes the main conclusions of this study and discusses future research directions. The appendices provide detailed derivations of the information bottleneck theory (\ref{app:B}), discussions on information emergence mechanisms in neural networks (\ref{app:C}), full proofs for i.i.d. Gaussian systems (\ref{app:D}), and extended proofs of the theorems (\ref{app:E}). 
\section{Related Work}
This section reviews the classical achievements and recent developments in the fields of high-dimensional data analysis and complex systems, with a focus on the breakdown of traditional metrics (distance concentration) in high-dimensional spaces, theoretical interpretations of emergence in complex systems, and recent explorations of information-theoretic approaches to these two core problems \cite{ref1,ref2}. 
Through a systematic review of the existing literature, we identify the theoretical gaps and challenges present in current research, thereby laying a theoretical foundation for the construction of the unified information-theoretic–statistical-mechanical framework proposed in this paper. 

\subsection{Metric Failure in High-Dimensional Data}
High-dimensional data analysis has long been challenged by the problem of concentration of distances.In early work, Beyer et al. first pointed out that nearest-neighbor search is severely affected by the curse of dimensionality in high-dimensional spaces. Subsequently, Aggarwal et al., through theoretical analysis and extensive empirical validation, demonstrated that as the dimensionality of data increases, traditional geometric metrics such as Euclidean distance and cosine similarity tend to converge, rendering them ineffective in distinguishing differences between samples \cite{ref17,ref24}.
In addition, research by Donoho and others further revealed the contradiction between the inherent sparsity of high-dimensional data and the phenomenon of distance concentration, indicating that although high dimensionality offers certain potential advantages in some contexts, it leads to serious performance bottlenecks in practical tasks such as clustering, dimensionality reduction, and nearest-neighbor retrieval \cite{ref24}. 
To address these issues, researchers have proposed a variety of manifold learning–based dimensionality reduction methods, such as ISOMAP, Diffusion Maps, t-SNE, and UMAP. These methods attempt to recover the low-dimensional structure of data through local neighborhood information. However, when dealing with extremely high-dimensional or complexly distributed data, traditional geometric approaches still face limitations in effectively capturing global dependencies \cite{ref30}. 
In recent years, some studies have introduced information-theoretic concepts such as mutual information and entropy to measure nonlinear dependencies among variables in data \cite{ref58, ref59}, in an attempt to provide a new theoretical explanation for the failure of metrics in high dimensions. However, most of these efforts remain empirical in nature and have yet to form a rigorous and unified theory.

\subsection{Emergence in Complex Systems}
Emergence in complex systems is a widespread phenomenon observed in physics \cite{ref3}, chemistry \cite{ref4}, biology \cite{ref13}, neuroscience \cite{ref6}, and the social sciences \cite{ref34}. It refers to the spontaneous appearance of novel macroscopic order or functional modules in systems composed of a large number of interacting microscopic units, typically under certain critical conditions. 
In statistical physics, phase transition theories such as the Ising model reveal how local coupling between spins and energy balancing lead to the formation of macroscopic ordered states near critical temperatures \cite{ref3,ref4}. 
These theories not only uncover long-range correlations and scale invariance in critical phenomena, but also provide a theoretical basis for explaining macroscopic order in other physical systems such as liquid crystals and superconductors \cite{ref7,ref9,ref14}. 
In the study of complex networks and graph theory, researchers have shown—through community detection, modular analysis, and topological data analysis—how local clustering of nodes can give rise to global functional modules. These findings have been validated in social networks, ecological systems, and neural networks \cite{ref34,ref35}. 
Meanwhile, in biology and neuroscience, phenomena such as cell differentiation, embryonic development, and the functional partitioning of neural systems are also regarded as classic examples of self-organized emergence under complex information transmission and local interactions\cite{ref60}. 
In social systems, local rules of individual decision-making and information diffusion can similarly give rise to macro-level phenomena such as collective behavior and opinion polarization\cite{ref61, ref62}. 
Despite the diversity of these studies, most models rely primarily on energy functions, local coupling, or network topology to explain emergent phenomena, while paying insufficient attention to information transmission, encoding capacity, and the quantitative characterization of emergence mechanisms \cite{ref7}. 
In recent years, some studies have attempted to employ tools such as the information bottleneck theory, mutual information estimation, and information-theoretic regularization to interpret emergence mechanisms. However, most of these efforts have focused on specific domains (e.g., representation learning in neural networks) or have relied on empirical approaches, and a rigorous cross-disciplinary theoretical framework has yet to be established. 

\subsection{A Unified Theoretical Framework from the Perspective of Information Theory}
Information theory, through the quantitative tools of mutual information and entropy, provides a universal framework for analyzing the intrinsic relationship between complexity and emergent behavior in systems. It reveals a dynamic unification between the curse of dimensionality and order formation under constraints of information encoding capacity \cite{ref46,ref47,ref48}. 
In recent years, with the widespread application of information-theoretic tools across disciplines \cite{ref36, ref40, ref46}, a growing body of work has begun to re-examine problems in high-dimensional data and complex systems through the lenses of mutual information, entropy, and the information bottleneck \cite{ref12, ref11}.
The information bottleneck theory proposed by Tishby et al. offered new insights into information compression in deep learning. Concurrently, several studies have attempted to apply Mutual Information Neural Estimation (MINE) and other modern methods to measure nonlinear dependencies in high-dimensional data. 
However, these studies typically focus on a single domain, such as representation learning in neural networks or dimensionality reduction in high-dimensional data, and lack a unified theoretical framework that connects the failure of high-dimensional metrics with emergence in complex systems \cite{ref11,ref12}. 
This study is developed precisely within this context: we propose the Information Dilution Theorem by defining a metric for mutual information efficiency and rigorously proving the mechanism by which the information-capturing capability of traditional geometric metrics degrades rapidly with increasing dimensionality in high-dimensional systems. 
In parallel, based on the concepts of interaction encoding capacity and spectral decomposition of the mutual information matrix, we construct the Emergence Critical Theorem, which quantitatively describes the inevitability of spontaneous emergence of new global features when a system's informational complexity exceeds the limit of low-dimensional encoding. 
Compared to existing work, the contribution of this paper lies in the construction of a unified theoretical framework that bridges high-dimensional data analysis and the study of emergence in complex systems. It offers a novel information-theoretic interpretation of their intrinsic connection and points toward future directions for algorithmic improvements and empirical validation \cite{ref2}. 
In summary, existing studies have achieved considerable progress in both high-dimensional data analysis and the investigation of emergent phenomena in complex systems, but these two lines of research have largely evolved in isolation. 
While traditional geometric methods and phase transition theories have explained metric failure and emergence phenomena respectively, they have not succeeded in establishing a cross-domain unified explanation from the perspective of information transmission and encoding capacity \cite{ref8,ref9}. 
From a cross-disciplinary perspective integrating information theory and statistical mechanics, this paper aims to fill this theoretical gap by constructing a unified framework that provides a quantitative basis for understanding information dilution in high-dimensional data and self-organized emergence in complex systems. 
We believe that this unified framework will offer a new theoretical foundation for areas such as high-dimensional data processing, deep learning, and complex system modeling. 
\section{Theoretical Foundations and Core Theorems}
\label{section3}
This section aims to clarify, from an information-theoretic perspective, that the two major phenomena—metric failure in high-dimensional spaces and emergence in complex systems—fundamentally stem from the same core contradiction: as the dimensionality of a system increases or its coupling structure becomes increasingly complex, the limited encoding capacity of the observer or interacting agent becomes insufficient to fully represent the system's vast informational structure. This insufficiency leads to the failure of traditional geometric metrics and compels the system to compensate for the lost information through the spontaneous generation of new global features—i.e., emergence. 
Based on this insight, the following sections will successively introduce the Information Dilution Theorem and the Emergence Critical Theorem. The former explains the inevitability of metric failure in high-dimensional systems via the decay of mutual information efficiency with increasing dimensionality, while the latter establishes rigorous conditions and criteria for the spontaneous emergence of new features within complex systems. Together, these two theorems form a unified explanatory framework for the above phenomena. 

\subsection{Problem Definition and the Metric of Mutual Information Efficiency} 
In many practical applications (such as gene expression, image features, and text embeddings), the system state can be represented as a vector composed of multiple random variables. 
\begin{equation}
S \;=\; (X_1,\,X_2,\,\dots,\,X_d). 
	\tag{1}
	\label{s=}
\end{equation}
If we assume that these random variables are mutually independent or weakly correlated, then due to the additivity of Shannon entropy, the total entropy of the system satisfies 
\begin{equation}
 H(S) \;=\; \sum_{i=1}^{d} H(X_i) \;=\; d \, H(X_1),  
	\tag{2}
\end{equation} 
This shows that the total information content of the system increases linearly with the dimension \( d \). 
On the other hand, consider a classical geometric metric \( D \), taking Euclidean distance as an example, defined as 
\begin{equation}
D \;=\; \|\,S \;-\; S^\prime\|, 
		\tag{3}
\end{equation}
where \( S^\prime \) is identically distributed and independent of \( S \). When the dimension \( d \) is large, the concentration of distances effect causes most sample pairs' distances to concentrate around a typical value, making the distribution of \( D \) highly peaked. 
As a result, the entropy of the distance distribution \( H(D) \) is constrained to the constant order \( O(1) \), which is far lower than the linear growth \( \Theta(d) \) of system entropy \( H(S) \) (see detailed derivation for i.i.d. Gaussian variables in Appendix~\ref{app:D}). 
To characterize the limitations of Euclidean distance in capturing information in high-dimensional systems, we define the mutual information efficiency metric as 
\begin{equation}
\eta(D) \;=\; \frac{I\bigl(D;\,S\bigr)}{H(S)}, 
\tag{4}
\label{N(d)}
\end{equation}
where \( I(D;S) \) denotes the mutual information between the metric \( D \) and the system state \( S \). This metric can be interpreted as the relative proportion of the system's original information retained by the metric \( D \). 
Since \( H(S) \) increases linearly with \( d \), while \( H(D) \) typically remains sublinear, \( \eta(D) \) approaches zero as \( d \) grows. 
This metric reveals the essence of "metric failure in high dimensions,” and also provides theoretical insight into how a system under encoding constraints may compensate for information loss through the emergence of new global features. 
\subsection{Information Dilution Theorem and Theoretical Analysis} 
Based on the above discussion, we propose and prove the "Information Dilution Theorem," which quantifies, from the perspective of mutual information, the significant degradation in the ability of Euclidean distance to capture system information as dimensionality increases. 
\vspace{0.25cm}

\textbf{Theorem 1 (Information Dilution Theorem)}\ $ $Let the system state \( S \) (defined in~\eqref{s=}) be a \( d \)-dimensional system of independent (or weakly correlated) random variables, and let \( D \) be any Euclidean distance or its weighted variant. As \( d \to \infty \), the mutual information efficiency of \( D \) with respect to \( S \) satisfies  
\begin{equation}
\eta(D) = \frac{I(D; S)}{H(S)} \leq O\left(\frac{1}{d}\right).
\tag{5}
\end{equation} 

This result indicates that although the total entropy \( H(S) \) increases linearly with \( d \), the actual amount of information extractable by the Euclidean distance \( D \), namely \( I(D;S) \), remains only at a constant or logarithmic level. Consequently, the mutual information efficiency \( \eta(D) \) decays at a rate of \( O\left(\frac{1}{d}\right) \). This theorem provides an information-theoretic explanation for the phenomenon of "distance failure" in high-dimensional spaces, and offers a theoretical foundation for understanding the performance degradation of many algorithms based on nearest-neighbor search or geometric assumptions (e.g., clustering, manifold learning) in high dimensions. The main proof strategy is as follows: 

\vspace{0.25cm}
First, for independent and identically distributed or weakly correlated random variables, we can write  
\begin{equation}
H(S) \;=\; \sum_{i=1}^d H(X_i) \;=\; O(d). 
\tag{6}
\end{equation} 
This implies that the system information grows proportionally (or approximately) with the dimensionality. 
Next, take the standard Euclidean distance as an example:  
\begin{equation}
D \;=\; \|\,S - S^\prime\|\;=\;\sqrt{\sum_{i=1}^d (X_i - X_i^\prime)^2},  
\tag{7}
\end{equation} 
where \( S^\prime \) is an independent and identically distributed copy of \( S \). Using the central limit theorem, it can be shown that \( D^2 \) approximately follows a chi-squared distribution with \( d \) degrees of freedom, with mean \( 2d\sigma^2 \) and standard deviation \( O(\sqrt{d}) \). After appropriate normalization (e.g., dividing by \( d \)), the mean of \( D \) remains constant, while the standard deviation shrinks at the rate of \( O(1/\sqrt{d}) \), resulting in a highly concentrated distribution for \( D \), and thus an entropy \( H(D) \) upper bounded by \( O(1) \). 
From an information-theoretic point of view, when the distribution of a random variable \( D \) becomes sharply peaked (i.e., has low entropy), its informational association with the system \( S \) is also limited. Formally, one can estimate using the data processing inequality\cite{ref40}: if \( D \) is concentrated in a narrow interval, then its entropy \( H(D) \) cannot reach  \( O(d) \)   , and may instead stay at  \( O(\log d) \) or even a constant order. Thus,  
\begin{equation}
I(D;S) \;\le\; H(D) \;\approx\; O(1) \quad\text{(or at least much less than order } d\text{)}.
\tag{8}
\end{equation} 
Therefore,  
\begin{equation}
\eta(D) \;=\; \frac{I(D;S)}{H(S)} \;\le\; \frac{O(1)}{O(d)} \;=\; O\Bigl(\frac{1}{d}\Bigr).
\tag{9}
\end{equation} 
This demonstrates that, in high-dimensional systems composed of independent variables, the mutual information efficiency of Euclidean distance inevitably decreases rapidly with dimension. 
In more general weakly correlated systems (satisfying \( \mathrm{Tr}(\Sigma_S) = O(d) \) and \( \lambda_{\max}(\Sigma_S) \leq \kappa \), where the total system variance grows linearly with dimension and no strongly coupled dominant modes exist), the extended derivation in Appendix~\ref{app:E} verifies that the upper bound \( \eta(D) \leq O\left(\frac{1}{d}\right) \) still holds. This result not only covers the case of independent Gaussian systems (Appendix~\ref{app:D}), but also ensures the applicability of the theorem to general weakly correlated systems through the spectral constraint \( \lambda_{\max}(\Sigma_S) \leq \kappa \). For detailed mathematical derivations, see Appendices~\ref{app:D} and~\ref{app:E}. 
In summary, Euclidean distance cannot match the linearly growing information of the system when the dimension \( d \) is large, and its mutual information efficiency asymptotically decays. This failure arises from the homogenization of pairwise distances in high-dimensional space and the insufficient entropy content of the metric itself. It reveals that the essence of the information dilution phenomenon lies in the \textbf{irreconcilable contradiction between the linear growth of system entropy and the sublinear growth of metric entropy}. This conclusion also lays the groundwork for introducing the "Emergence Critical Theorem": when the internal informational patterns of a system far exceed the dimensional capacity of an observer or interacting agent, new global features will spontaneously emerge to compensate for the information loss incurred by conventional metrics. Section~3.3 will discuss this issue in detail. 

\subsection{Emergence Critical Theorem} 
\subsubsection{Definitions} 
In the previous section, we revealed through the "Information Dilution Theorem" that in high-dimensional systems, although the total entropy \( H(S) \) increases linearly with dimension, the entropy growth of traditional low-dimensional geometric metrics significantly lags behind, making them incapable of capturing the full information of the system. 
Meanwhile, as the dimensionality increases, the modes of interdependence among variables in the system become increasingly diverse and complex, thereby increasing the number of independent information interaction patterns. 
In other words, high-dimensional spaces not only pose challenges such as "concentration of distances" for geometric metrics, but also conceal rich and non-redundant internal informational structures. 
When such internal informational complexity exceeds the limited encoding capacity of the observer (or more generally, the interacting agent), the system will spontaneously generate new global features to supplement the information lost in the low-dimensional representation. 
To quantitatively describe this phenomenon, we introduce two key concepts: \textbf{information structural complexity} and \textbf{interaction encoding capacity}, which serve as the theoretical bridge connecting Theorem 1 and Theorem 2. 
\begin{enumerate}
\item \textbf{Information Structural Complexity:} 
The total information \( H(S) \) reflects only the aggregate uncertainty of the variables, but it cannot reveal how these variables interact in diverse, independent, and non-redundant ways to produce complex dependencies. 
To capture this internal structure, we introduce the notion of "information structural complexity," which quantifies the effective number of independent information interaction patterns in the system. 
Specifically, we construct the mutual information matrix  
\begin{equation}
I \in \mathbb{R}^{d\times d}, \quad I_{ij} = I(X_i;X_j), \quad 1\le i,j\le d,
\tag{10}
\label{I}
\end{equation} 
where \( I(X_i;X_j) \) measures not only the linear correlation between \( X_i \) and \( X_j \), but also captures their nonlinear dependencies. 
This matrix provides a comprehensive representation of the strength and structure of information exchange among system variables. 
Intuitively, if the rank \( \mathrm{Rank}(I) \) is high, it indicates the existence of diverse and independent information interaction patterns in the system; conversely, if the rank is low, the system's information is concentrated in a few dominant modes, and the overall dependency is relatively simple. 
To quantify this characteristic more precisely, we perform singular value decomposition (SVD) on the mutual information matrix \( I \). Since \( I \) is symmetric, the decomposition can be written as  
\begin{equation}
I = U\Lambda U^\top, \quad \Lambda = \mathrm{diag}(\sigma_1,\,\sigma_2,\,\dots,\,\sigma_r),
\tag{11}
\end{equation} 
Here, \( r = \mathrm{Rank}(I) \) denotes the rank of matrix \( I \), and the singular values satisfy \( \sigma_1 \ge \sigma_2 \ge \cdots \ge \sigma_r > 0 \). 

To avoid treating extremely small singular values—arising from noise or weak dependencies—as meaningful patterns, we introduce the normalized weights 
\begin{equation}
p_i = \frac{\sigma_i}{\sum_{j=1}^{r}\sigma_j},\quad i=1,\dots,r.
\tag{12}
\end{equation}

Using this probability distribution, we define the information structural complexity (or effective rank) of the system as 
\begin{equation}
\mathcal{C}(S) \;=\; \mathrm{erank}(I) \;=\; \exp\!\Biggl(-\sum_{i=1}^{r} p_i \ln p_i\Biggr).
\tag{13}
\end{equation}
When all singular values are equal, each \( p_i = 1/r \), yielding \( \mathrm{erank}(I) = r \). When only a few singular values dominate and the rest are significantly smaller, then \( \mathrm{erank}(I) \) becomes significantly smaller than \( r \). 
Therefore, \( \mathcal{C}(S) \), as a dimensionless index, not only reflects the total uncertainty of system variables, but more importantly, robustly identifies the number of independent modes that play key roles in information interaction, thus quantifying the system's intrinsic informational complexity. 
From a physical perspective, this index provides a concept similar to "degrees of freedom" or "rank," but with stronger discriminative power. 
It eliminates redundant modes introduced by noise or weak dependencies and effectively captures the dominant coupling structures that have a decisive influence on global system behavior. 
In particular, this metric can also be regarded as a tool for \textbf{detecting the non-uniformity in the distribution of degrees of freedom}. 
This tool has potential applications in various complex systems such as biological networks, social networks, and deep neural networks, because in these systems, only those information modes with sufficient "energy" or "weight" can drive global phase transitions or self-organization. 
By introducing the concept of information structural complexity, we establish a quantitative foundation for the later discussion on the critical relationship between interaction encoding capacity and system-internal complexity. 
\item \textbf{Interaction Encoding Capacity} 
In traditional information processing frameworks, the "observer" is usually regarded as an external and passive recipient of information, whose limited encoding capacity restricts the ability to fully capture all the information within a high-dimensional system. 
Consider that the system state \( S \) (as defined in~\eqref{s=}) is compressed into a low-dimensional representation via a mapping \( f \). Due to physical constraints (such as the number of sensors, number of principal components, or the dimension of hidden layers in neural networks), we have 
\begin{equation}
\dim(f) \le k.
\tag{14}
\end{equation}
Under this constraint, the mutual information \( I(f;S) \) between the mapping \( f \) and the original system \( S \) quantifies how much information can be retained during compression. 
Based on this, we define the interaction encoding capacity as 
\begin{equation}
C = \max_{f:\,\dim(f)\le k} I(f;S).
\tag{15}
\end{equation}
This definition represents the maximum mutual information achievable among all mappings that satisfy the low-dimensional constraint. It is the theoretical upper bound of how much of the system's information the observer can "capture" or "represent" under limited encoding resources. 
This concept provides a quantitative basis for later discussions on how insufficient encoding resources lead to information loss and the spontaneous emergence of new features. 
Furthermore, to better reflect the interactions among all information processing entities in real systems, we extend the definition of the "observer" to a more general interacting object. 
Under this extension, whether it is an external observer or an internal subsystem (such as a neural network layer or a subgroup in an ecosystem), each is assigned a limited encoding capacity \( C \). 
However, it is important to note that \( C \) is measured in bits (information quantity), whereas the previously defined information structural complexity \( \mathcal{C}(S) \) describes the number of independent informational modes in the system. These two quantities differ in dimension. 
To address this mismatch, we introduce a dimensional transformation to convert the encoding capacity \( C \) into an equivalent number of independent informational modes. 
Specifically, we perform singular value decomposition on the mutual information matrix between system variables, obtaining the singular value sequence 
\begin{equation}
\{\sigma_1, \sigma_2, \dots, \sigma_r\} \quad (\sigma_1\ge\sigma_2\ge \cdots\ge \sigma_r>0),
\tag{16}
\end{equation}
We then define the transformed encoding capacity as 
\begin{equation}
C^{\prime} = \max \left\{ n \in \mathbb{N} \,\Big|\, \sum_{i=1}^{n} \sigma_i \leq C \right\}.
\tag{17}
\label{17}
\end{equation}
In this definition, $C'$ represents the maximum number of independent information modes that the observer (or interacting agent) can capture under limited encoding resources $C$. 
This unit conversion brings the upper limit of encoding capacity, originally measured in information quantity, to the same scale as the number of independent information modes in the system (e.g., the rank or effective rank of the mutual information matrix), thus providing a mathematical basis for the criterion: "when the system's information structural complexity exceeds the encoding upper bound, new global features spontaneously emerge." 
In summary, we introduced two key concepts: information structural complexity $\mathcal{C}(S)$ and interaction encoding capacity $C$, along with the converted encoding capacity $C'$. 
The former quantifies the number of non-redundant and independent information interaction modes in the system via singular value decomposition of the mutual information matrix and normalized weights; 
the latter describes the maximum amount of information that an information-processing entity can capture under low-dimensional mapping constraints, and by unit conversion, it becomes comparable to the information structural complexity. 
To better understand the meaning of $C'$, we interpret it in an "information budget" framework. 
The total encoding capacity $C$ is treated as the total available information budget. 
The singular values \(\sigma_i\) of the mutual information matrix can be regarded as the "cost" or "information resource consumption" required to encode or capture the $i$-th independent information interaction mode (sorted by importance). 
Encoding a stronger or more informative mode (with larger $\sigma_i$) naturally requires a higher cost. 
Thus, $C'$ defined in (\ref{17}) intuitively characterizes the maximum number of independent information modes (ordered by importance) that the observer or interaction unit can "afford"(i.e., fully encode or effectively capture) within the given information budget $C$. 
This number $C'$ therefore serves as an upper bound on the system's intrinsic information dimension that can be effectively perceived under encoding limitations. 
In subsequent sections, we will construct rigorous mathematical criteria based on these two quantitative indicators, proving that when the intrinsic information complexity of the system exceeds the converted encoding capacity $C'$, traditional low-dimensional representations inevitably exhibit information loss, thereby triggering the spontaneous emergence of new global features. 
\end{enumerate}
\subsubsection{Theorem Statement and Proof} 
\textbf{Theorem 2 (Emergence Critical Theorem)} 
Let the mutual information matrix \( I \) of the system state \( S \) (defined in~\eqref{s=}) have an effective rank (i.e., information structural complexity) denoted by \( \mathcal{C}(S) \), and let the interaction agent's encoding capacity after dimensional transformation be bounded above by \( C^{\prime} \). 
If \(\mathcal{C}(S) > C^{\prime}\)
then there must exist at least one new global feature \( f_{\mathrm{new}} \), such that the mutual information between \( f_{\mathrm{new}} \) and the system state \( S \) satisfies  
\begin{equation}
\frac{I(f_{\mathrm{new}};S)}{H(S)} > \theta,
\tag{18}
\label{>0}
\end{equation} 
where \( \theta \in (0,1) \) is a predefined significance threshold (e.g., \( \theta = 0.05 \)). 
In other words, when the number of independent informational modes in a system exceeds the encoding capacity, the existing "low-dimensional view" is insufficient to capture all the information, forcing the system to generate a new nonlinear feature \( f_{\mathrm{new}} \) to supplement its representation. This results in the appearance of new macroscopic order parameters or functional modules. 
This answers the question: "Why do new, unforeseen macroscopic features appear as the system's information dimension increases?"—because the system requires new global quantities to describe hidden patterns. This is the essence of "emergence" in many complex systems. 
To fully demonstrate the validity of this theorem, we divide the proof into four major parts. These will sequentially reveal the internal relationship between the insufficiency of low-dimensional representation and the inevitable generation of new global features, when the system's internal informational dimension exceeds the observer's encoding capacity: 
\begin{enumerate}
\item \textbf{Limitations of Low-Dimensional Encoding} 
Consider the system state \( S \) (defined in~\eqref{s=}) and construct the mutual information matrix \( I \) as defined in~\eqref{I}. According to the previous definitions, the system's information structural complexity \( \mathcal{C}(S) \) reflects the effective rank of \( I \), i.e., the number of non-redundant and independent informational modes within the system. 
To assess the significance of these informational modes, we perform singular value decomposition on \( I \): 
\begin{equation}
I = U \Lambda U^\top,
\tag{19}
\end{equation}
where \( U \) is an orthogonal matrix, and \( \Lambda = \mathrm{diag}(\sigma_1, \sigma_2, \dots, \sigma_r) \) is a diagonal matrix containing non-negative singular values, with \( r = \mathrm{Rank}(I) \) being the rank of \( I \). 
In this decomposition, each singular value \( \sigma_i \) represents the "strength" or contribution of the corresponding informational mode. 
When the observer (or more generally, the interacting agent) encodes the system, suppose its transformed encoding capacity is bounded above by \( C^{\prime} \), i.e., the agent can construct at most \( C^{\prime} \) low-dimensional features to capture system information. 
When \( \mathcal{C}(S) \le C^{\prime} \), the major informational interaction modes in the system can be sufficiently represented within the available encoding dimension, and the low-dimensional representation can adequately reconstruct the system's dependency structure. 
However, when \( \mathcal{C}(S) > C^{\prime} \), there inevitably exist some informational modes—especially those corresponding to smaller singular values—that cannot be embedded into the limited \( C^{\prime} \)-dimensional encoding space. 
It should be noted that these small singular values may, in practice, correspond to weak correlations or noise. However, when they are numerous, even if each one contributes only a little, collectively they constitute a non-negligible amount of missing information. 
This "tail" of uncaptured information reflects the inadequacy of the low-dimensional representation in expressing the system's complex informational structure, and thereby provides the driving force for the system to spontaneously generate new global features \( f_{\mathrm{new}} \). 
The purpose of this new feature is to compensate for the encoding gap. 
Thus, through spectral decomposition of the mutual information matrix, we intuitively illustrate that when the system's internal informational structural complexity exceeds the encoding capacity upper bound \( C^{\prime} \), information loss in low-dimensional representations becomes inevitable, which drives the emergence of new global features. 
\item \textbf{Capturing Residual Information via the Information Bottleneck Principle} 
To capture the important residual information in the system that is not sufficiently encoded by traditional low-dimensional representations, we adopt the Information Bottleneck (IB) theory to model the encoding process. 
Let \( f \) be a compressed representation of the system state \( S \), whose dimension is constrained by physical limits, i.e., \( \dim(f) \le k \), corresponding to an encoding capacity upper bound \( C \). 
Under this condition, we consider the following constrained optimization problem: 
\begin{equation}
\max_{f} \, I(f; S) \quad \text{subject to} \quad H(f) \le C.
\tag{20}
\end{equation}
Here, \( I(f; S) \) denotes the amount of information that the compressed representation \( f \) retains about the system state \( S \), while \( H(f) \) quantifies the total information contained in \( f \). 
The goal is to maximize the amount of relevant information \( f \) can capture from \( S \), under the constraint that its total information content does not exceed the encoding budget \( C \). 
When the number of independent informational modes in the system exceeds the encoding capacity, traditional linear mappings (i.e., "old features") often fail to achieve this optimal objective, since they only capture the most dominant statistical dependencies in the system, and cannot reflect weaker or more complex interaction patterns. 
In other words, linear low-dimensional mappings have inherent limitations when dealing with residual information beyond the encoding capacity. 
To overcome this limitation, the system must introduce nonlinear mappings to generate a new feature \( f_{\mathrm{new}} \). 
This new feature is specifically designed to capture residual information that cannot be represented via traditional low-dimensional linear combinations, thereby compensating for the shortcomings of the original encoding and improving the overall information capture. 
Through this process, the introduction of \( f_{\mathrm{new}} \) enables the mutual information between the system and its representation to exceed the pre-defined threshold~\eqref{>0}. 
From a physical perspective, limited encoding resources inherently lead to the loss of part of the system's statistical dependency structure when compressing high-dimensional information. 
Low-dimensional representations can only capture the most significant information modes in the system, while weaker or more intricate dependencies are left unrepresented. 
This "information gap" is precisely the bottleneck effect in the encoding process. 
Information Bottleneck theory formalizes this distortion phenomenon by maximizing target information transmission under entropy constraints. 
When the system's intrinsic informational modes exceed the encoding limit, the original mapping cannot retain all key information. As a result, the system inevitably needs to generate new nonlinear features to compensate for this deficiency and achieve a more complete representation of the global information. 
This self-organizing adjustment mechanism reflects a necessary strategy adopted by the system to maintain its internal complexity under resource constraints, and serves as the physical basis for the phenomenon of emergence. 
\item \textbf{Lagrangian Optimization and Distortion-Driven Feature Generation} 
To formally describe the information loss that arises when an observer (or interacting agent) compresses system information under limited encoding capacity, we employ the Information Bottleneck (IB) principle to construct a Lagrangian optimization framework, incorporating the encoding constraint into the objective function. 
Let \( f \) be a compressed representation of the system state \( S \), and let \( Y \) denote a target variable associated with the emergent feature of the system. 
We aim to ensure that \( f \) retains as much information as possible about \( Y \), under the constraint that the encoding budget \( C \) is not exceeded. To this end, we construct the following Lagrangian: 
\begin{equation}
\mathcal{L} = I(f; Y) - \lambda \Bigl( I(S; f) - C \Bigr),
\tag{21}
\end{equation}
Here, \( I(f; Y) \) quantifies the amount of information transmitted between \( f \) and the target variable \( Y \), while \( I(S; f) \) denotes the total amount of information that \( f \) extracts from \( S \). The parameter \( \lambda \ge 0 \) is a Lagrange multiplier balancing the trade-off between information preservation and the encoding resource constraint. 
By variationally solving the conditional distribution \( p(f \mid S) \), we obtain the optimal mapping in the following form: 
\begin{equation}
p(f \mid S) \propto p(f) \exp\!\left(-\frac{1}{\lambda} D_{\mathrm{KL}}\Bigl(p(Y \mid S) \,\Vert\, p(Y \mid f)\Bigr)\right).
\tag{22}
\end{equation}
This expression shows that the information transfer is more effective when \( f \) is able to make \( p(Y \mid f) \) closely approximate \( p(Y \mid S) \); however, under the constraint of limited encoding capacity \( C \), if the original low-dimensional mapping \( f \) fails to fully reconstruct all dependencies between \( S \) and \( Y \), then the KL divergence \( D_{\mathrm{KL}}\bigl(p(Y \mid S)\,\Vert\, p(Y \mid f)\bigr) \) will deviate significantly from zero, indicating a high degree of encoding distortion. 
This distortion is the intrinsic driving force that compels the system to generate a new feature \( f_{\mathrm{new}} \). 
By employing a nonlinear mapping, the new feature aims to compensate for the residual information that traditional low-dimensional representations fail to capture under limited resources, thereby enabling the mutual information between the new feature and \( S \) to reach the preset threshold~\eqref{>0}. 
Physically, this process resembles how constrained systems, under energy or free energy limitations, spontaneously generate new structures through self-organization to compensate for local informational insufficiency. 
Thus, the inability of the original mapping to capture the full complexity of the system becomes the driving force for the emergence of new global features. When the number of independent informational modes in the system exceeds the available encoding capacity, additional nonlinear features must be introduced to express the global dependency structure completely. 
This mechanism reveals the physical essence of emergence: a self-regulatory process whereby the system seeks to maintain its overall informational expressiveness. 
Therefore, the Lagrangian optimization framework not only quantifies the distortion induced by limited encoding resources, but also reveals that when the number of intrinsic information modes exceeds the capacity, the inadequacy of the original mapping necessarily leads the system to adopt new nonlinear mappings to generate additional features for a more complete global representation. 
For detailed derivations and mathematical tools related to the Lagrangian optimization in Information Bottleneck theory, see Appendix~\ref{app:B}. 
\item \textbf{Generalized Case} 
In more general scenarios, if the system's information modes are primarily characterized by nonlinear dependencies, linear mappings often fail to capture the entire structure. 
In such cases, kernel methods can be employed to construct nonlinear features: 
\begin{equation}
f_{\mathrm{new}}^{\mathrm{nonlin}} = \phi(S)^\top \beta,
\tag{23}
\end{equation}
where \( \phi \) is a kernel mapping function that projects high-dimensional data into a Reproducing Kernel Hilbert Space (RKHS), and \( \beta \) is a weight vector. 
Under the condition \( \mathcal{C}(S) > C^{\prime} \), the mutual information of this nonlinear feature, \( I(f_{\mathrm{new}}^{\mathrm{nonlin}}; S) \), will also exceed the predefined threshold \( \theta \), thus providing a rigorous mathematical criterion for phenomena such as phase transitions, gene differentiation, and other types of emergence. 
For further discussion, see Appendix~\ref{app:E}, where kernel-based spectral decomposition of the mutual information matrix and feature construction methods are explored in more depth. 
In conclusion, Theorem 2 clearly shows that when the system's information structural complexity \( \mathcal{C}(S) \) exceeds the limited encoding capacity \( C \) of the interacting agent, traditional low-dimensional mappings inevitably fail to fully capture all of the system's information, resulting in the omission of certain key informational components. 
As a consequence, the system spontaneously generates new nonlinear features \( f_{\mathrm{new}} \), such that their relative contribution to the overall information representation satisfies\eqref{>0}
This mechanism not only answers the question of why entirely new macroscopic features tend to emerge as the system's informational dimensionality increases, but also provides a rigorous theoretical foundation for understanding the intrinsic driving forces behind emergence in complex systems. 
Moreover, it lays the groundwork for future applications and empirical validations in statistical physics, biological systems, and deep neural networks. 
\end{enumerate}	
\subsection{Physical Interpretation} 
In high-dimensional systems, as the system dimensionality increases, the total amount of microscopic information expands rapidly. 
However, traditional low-dimensional metric tools (such as Euclidean distance) often fail to capture such complex informational structures. 
According to the Information Dilution Theorem, when the system is composed of a large number of independent or weakly correlated variables, its total entropy \( H(S) \) grows linearly, whereas the information captured by low-dimensional metric tools grows only at constant or logarithmic rate. 
From a physical perspective, this mismatch resembles the phenomenon in statistical physics where the random fluctuations of individual particles, when averaged macroscopically, leave behind only a limited number of thermodynamic variables: the complex local fluctuations become "diluted" at the global scale, making traditional measures incapable of distinguishing between different states. 
Meanwhile, the Emergence Critical Theorem reveals from another angle how insufficient encoding capacity leads to self-organization. 
When the informational structural complexity formed by mutual information among system variables exceeds the encoding capacity of the observer or interacting agent, a "gap" inevitably appears in the original low-dimensional feature representation. 
This gap resembles the critical point in phase transitions: when the complex interactions among local degrees of freedom can no longer be fully captured by existing statistical descriptions, the system undergoes a sudden transition and spontaneously generates new global order parameters. 
For instance, in ferromagnetic phase transitions, the previously disordered local spins align rapidly into a globally magnetized state upon reaching the critical temperature. 
Similarly, under information bottleneck constraints, the newly generated global feature can be regarded as a "re-encoding" of mutual information that is not captured by the original low-dimensional representation, thereby partially recovering the completeness of the system's information expression. 
Overall, these two theorems together form a unified explanatory framework: when microscopic informational complexity greatly exceeds the limited encoding resources of the observer or interaction object, traditional local metrics will inevitably "fail", and the system must generate new global variables through self-organization to compensate. 
This viewpoint not only aligns well with phenomena such as phase transitions, self-organization, and emergence in physics, but also provides a theoretical foundation for understanding real-world phenomena such as spontaneous formation of hierarchical features in deep neural networks, the appearance of critical nodes in gene expression networks during cell differentiation, and sudden transitions in group behaviors within complex social systems. 
Through this interdisciplinary perspective that combines information theory and statistical mechanics, we gain a deeper understanding that the "order" in nature is not pre-designed, but rather the result of spontaneous emergence arising from a dynamic tension between an overabundance of microscopic degrees of freedom and the limited representational capacity of the observer or interacting agent. 
The emergence condition can be precisely characterized via the quantitative relationship between information structural complexity and encoding capacity. 

\section{Applicability and Implications of the Theory} 
\label{section4}

In Section~\ref{section3}, we introduced the Information Dilution Theorem and the Emergence Critical Theorem, which from an interdisciplinary perspective combining information theory and statistical mechanics, explained the intrinsic mechanisms behind the failure of traditional low-dimensional metrics in high-dimensional systems and the spontaneous emergence of new global features. 

Building on these theoretical results, this section further explores their applicability in real-world complex systems, and proposes corresponding experimental designs and future research directions, thereby offering insights for theoretical validation and interdisciplinary applications. 

\subsection{Theoretical Applicability and Validation Framework} 
Existing studies have shown that in real systems such as phase transitions in the Ising model and biological developmental differentiation, mutual information and correlation tend to increase significantly near the critical point~\cite{ref7, ref31}, reflecting a rise in the number of independent informational modes. 
This provides indirect support for the Emergence Critical Theorem proposed in this paper. 
Based on this theoretical framework, we propose verification approaches and application ideas for the following types of experiments and numerical simulations. 
First, in the context of \textbf{Ising model numerical simulations}, the Ising model serves as a classical example of phase transition and exhibits clear critical behavior. 
Specifically, one can collect large amounts of spin configuration data at different temperatures, compute the mutual information \( I(X_i; X_j) \) between arbitrary spin variables \( X_i \) and \( X_j \), and construct the mutual information matrix \( I \). 
Then, by performing spectral decomposition, one can analyze how its effective rank \( \mathcal{C}(S) \) varies with temperature and system size. 
If a significant jump in \( \mathcal{C}(S) \) is observed near the critical temperature, and this jump exceeds a predefined encoding capacity threshold \( C \), then it would directly validate the predictions of the Emergence Critical Theorem. 
Second, in the context of \textbf{high-dimensional gene expression data analysis}, single-cell RNA sequencing (scRNA-seq) technologies generate gene expression profiles that typically span thousands to tens of thousands of dimensions. 
Traditional Euclidean distance metrics often fail in such high-dimensional settings~\cite{ref23, ref35}, while using mutual information to quantify dependencies among genes can better capture nonlinear or complex interactions. 
By constructing a mutual information matrix from gene expression data and performing spectral analysis, one can evaluate the intrinsic informational structural complexity of the system. 
When a sudden increase in \( \mathcal{C}(S) \) is observed during a specific developmental or differentiation stage, it may signify the emergence of new modules or functional groups within the gene regulatory network, 
This provides a novel information-theoretic explanation for phase transition or mutation mechanisms in biological systems~\cite{ref18, ref29}. 
In addition, for \textbf{dynamical and non-equilibrium systems}, such as opinion propagation in social networks or spatiotemporal activity in neural systems, the system state evolves over time. 
In this case, one can construct a time-dependent mutual information matrix \( I(t) \), and track the evolution of its effective rank \( \mathcal{C}(S(t)) \), in order to detect potential critical transitions during the dynamic process\cite{ref22, ref40}. 
If a significant transition in \( \mathcal{C}(S(t)) \) is observed at a certain moment, and this corresponds closely to a macroscopic system change (e.g., opinion polarization or neural activation pattern switching), it would further support the applicability of the Emergence Critical Theorem in dynamical systems. 
This provides a new theoretical perspective for predicting and controlling non-equilibrium systems. 
In the above experimental designs, practical issues such as data noise, sparsity, and high-dimensional computational complexity must be carefully considered. 
To address these, one may apply techniques such as randomized SVD~\cite{ref27}, kernel methods~\cite{ref37}, or graphical models~\cite{ref33} for efficient approximation of the mutual information matrix. 
Moreover, synthetic datasets can be constructed to first verify the \( O(1/d) \) information dilution effect in systems composed of high-dimensional independent or weakly correlated variables, thereby offering preliminary theoretical support for the experimental framework. 
The above validation schemes are illustrative examples rather than completed implementations. Future research may develop experimental or numerical studies under these scenarios to more comprehensively test and refine the applicability of the Information Dilution Theorem and the Emergence Critical Theorem in high-dimensional and complex systems. 
\subsection{Cross-Disciplinary Applications} 
\subsubsection{UMAP+ and Persistent Mutual Information Homology} 
Traditional manifold learning methods (such as t-SNE and UMAP) primarily rely on local geometric similarity to achieve low-dimensional embedding. However, in high-dimensional spaces, due to the information dilution effect, conventional geometric metrics such as Euclidean distance often fail to preserve the global structure of the data\cite{ref35}. 
This limitation not only reduces the discriminative power of the low-dimensional representation, but also undermines the stability of clustering and classification results that are based on geometric distances\cite{ref23,ref29}. 
To more comprehensively capture the intrinsic statistical dependencies and multiscale structures of data~\cite{ref18}, we propose two improvements: one is to incorporate mutual information-based weights into manifold learning, and the other is to leverage topological data analysis tools to reveal global informational patterns and relate them to the Emergence Critical Theorem. 
First, we propose the idea of \textbf{UMAP+}. In standard UMAP, the low-dimensional embedding is typically optimized by minimizing the following objective function: 
\begin{equation}
\min_{f} \sum_{(i,j)} \|f(x_i) - f(x_j)\|^2,
\tag{24}
\end{equation}
This objective mainly depends on local Euclidean distances. However, in high-dimensional settings, Euclidean distances tend to become uniform due to information dilution, and thus fail to reflect the true statistical dependencies between variables. 
To address this issue, the UMAP+ approach introduces mutual information-based weights into the original objective, modifying it as follows: 
\begin{equation}
\min_{f} \sum_{(i,j)} w_{ij} \cdot \|f(x_i) - f(x_j)\|^2, \quad \text{where } w_{ij} = \Phi\!\bigl(I(X_i; X_j)\bigr).
\tag{25}
\end{equation}
Here, \( I(X_i; X_j) \) denotes the mutual information between variables \( X_i \) and \( X_j \), and the function \( \Phi \) transforms mutual information into positive weights, such that point pairs with higher statistical dependence receive greater emphasis during embedding. 
This approach aims to compensate for the loss of discriminative power of traditional Euclidean distances in high-dimensional settings, thus improving the capture of global structure and key functional information. 
By incorporating mutual information weights, the UMAP+ framework not only possesses strong theoretical motivation, but also provides a practical tool grounded in the theories of information dilution and emergence. 
Secondly, we introduce the \textbf{Persistent Mutual Information Homology} framework based on information topological analysis. 
Topological Data Analysis (TDA), particularly persistent homology, enables the extraction of global topological features (e.g., connected components, loops, voids) across multiple scales. 
We propose constructing an information topology using the mutual information matrix. Let \( I_\epsilon \) denote the matrix obtained by thresholding the original mutual information matrix such that entries less than \( \epsilon \) are set to zero. 
By varying \( \epsilon \), a filtration is constructed. We define the topological scale of the information network as: 
\begin{equation}
PH_{\mathrm{info}}(S) = \Bigl\{(b_i,d_i) \,\Big|\, b_i = \inf\{\epsilon \mid \mathrm{Rank}(I_\epsilon)=i\},\; d_i = \sup\{\epsilon \mid \mathrm{Rank}(I_\epsilon)=i\}\Bigr\}.
\tag{26}
\end{equation}
In this framework, long-lived topological features (e.g., persistent loops or cavities) reflect stable informational patterns within the data. 
Theoretically, if the number of such persistent features in \( PH_{\mathrm{info}}(S) \) exceeds the interaction encoding capacity \( C \), it indicates that traditional low-dimensional representations are insufficient to capture all independent information modes in the system. 
This matches the condition described by the Emergence Critical Theorem—namely, the system must spontaneously generate new global features to fill this "encoding gap." 
Therefore, the persistent mutual information homology approach not only provides a quantitative tool for analyzing multiscale data structures, but also serves as a bridge between theory and implementation for understanding the emergence of new functional modules. 
Together, the UMAP+ and persistent mutual information homology approaches aim to mitigate the limitations brought by information dilution in high-dimensional data—one by enhancing global structural fidelity in embedding, and the other by revealing intrinsic multiscale topological features. 
These methods provide a theoretical foundation for functional module identification and dynamic behavior analysis, and offer guidance for cross-disciplinary applications of the Emergence Critical Theorem. 
\subsubsection{Information-Theoretic Mechanism of Hierarchical Feature Emergence in Neural Networks} 
In artificial neural networks, feature representations across layers form a progressively hierarchical structure. This formation can be fundamentally regarded as an instance of "informational emergence." 
Based on the theoretical framework proposed in this work, we interpret the representation learning process in deep neural networks through the lens of the Information Dilution Theorem and the Emergence Critical Theorem, leading to the following conclusions: 
\begin{enumerate}
    \item Each layer of a neural network inevitably undergoes a transition from information compression to new feature emergence when processing high-dimensional inputs~\cite{ref23, ref11}. 
    \item The ability of deep networks to spontaneously learn high-level abstract features is not accidental, but rather a necessary consequence of limited encoding capacity \( C \) at each layer from an information-theoretic perspective~\cite{ref12, ref7}. 
\end{enumerate}
First, each layer in a neural network must compress its high-dimensional input in order to extract relevant features~\cite{ref23, ref29}. 
Since the encoding capacity of each layer is constrained by factors such as the number of neurons, the nonlinearity of activation functions, and other physical limitations~\cite{ref11, ref12}, this capacity—denoted as \( C \)—may be insufficient when the input information complexity (measured by the effective rank of the mutual information matrix) exceeds it. 
As a result, traditional low-dimensional mappings become incapable of representing all statistical dependencies present in the input~\cite{ref4, ref23}. 
To compensate for this information loss, the network necessarily generates new feature representations through nonlinear transformations—effectively "re-encoding" the residual information. 
This phenomenon is consistent with the Emergence Critical Theorem: when \( \mathcal{C}(S) > C^{\prime} \), new global features \( f_{\mathrm{new}} \) will emerge spontaneously, contributing mutual information satisfying~(\ref{>0}). 
Furthermore, from the perspective of layered architecture, lower layers typically capture local, low-level features, whereas higher layers, through multiple nonlinear transformations, extract more abstract representations. 
When the output information complexity of a given layer exceeds its encoding capacity, the network must generate new global representations to adequately describe the remaining information. This process is a direct manifestation of information emergence. 
Hence, the spontaneous emergence of high-level abstract features in deep networks not only explains the mechanism behind hierarchical representation formation, but also offers an information-theoretic interpretation of their "black-box" behavior. 
In practice, this information-theoretic mechanism can guide neural network design. For example, one may introduce mutual information regularization terms or apply information bottleneck constraints at each layer~\cite{ref12, ref1}, thereby encouraging the network to capture non-redundant information within its limited encoding capacity. 
Experimentally, one can analyze the rank of mutual information matrices at each layer and track their evolution during training. If the effective rank exceeds the encoding capacity \( C \), the emergence of new features can then be correlated with improvements in model performance. 
This provides both a theoretical foundation and an empirical direction for the design of interpretable and efficient deep neural networks. (See Appendix~\ref{app:C} for details.) 
\subsection{Future Works} 
In summary, this study has explored the potential of the Information Dilution Theorem and the Emergence Critical Theorem in high-dimensional complex systems from three perspectives: theoretical applicability, experimental validation, and cross-domain applications. 
The proposed theoretical framework has already received indirect empirical support in the contexts of the Ising model~\cite{ref2} and gene expression data~\cite{ref32}. It also offers new perspectives for improving manifold learning algorithms and interpreting hierarchical representations in deep neural networks. 
Future research can further refine and expand this theoretical framework along the following directions: 
\begin{enumerate}
    \item \textbf{Applicability under complex coupling conditions:} 
    The current derivations are based primarily on assumptions of independent or weakly correlated variables. Future studies should explore how strong coupling or block-wise coupling affects the behavior of information dilution and emergence, and determine whether the theorems remain valid or require modification to accommodate more intricate interaction structures. 
    \item \textbf{Efficient computation of mutual information matrix rank in high dimensions:} 
    For extremely high-dimensional data (e.g., genomic datasets or brain connectomes), directly computing the mutual information matrix and its rank is computationally intensive and sensitive to noise. Future efforts should focus on developing efficient algorithms using randomized SVD, kernel-based approaches, or graphical models to accurately estimate mutual information matrix ranks in large-scale datasets, thereby supporting both theoretical validation and real-world applications. 
    \item \textbf{Cross-disciplinary applications in deep learning:} 
    Neural networks, as multi-layered information processing systems, are intrinsically related to the Emergence Critical Theorem through their spontaneous generation of hierarchical features. Future work may involve incorporating mutual information regularization or bottleneck constraints within deep networks, analyzing the rank of mutual information matrices in hidden layers, and verifying the link between emergent features and network performance. 
    Furthermore, by applying this theory to real-world tasks in image processing, natural language understanding, and beyond, researchers can assess its practical interpretability and efficiency, thereby providing a theoretical foundation for the design of more transparent and effective models. 
    \item \textbf{Temporal dynamics in non-equilibrium systems:} 
    In dynamic systems such as social networks and neural activity, the rank of the mutual information matrix may evolve over time, reflecting potential phase transitions or sudden changes in system state. 
    Future research can leverage time-series data to study the spatiotemporal evolution of mutual information ranks and investigate whether such transitions correlate with changes in functional system states, thereby offering new theoretical tools for the prediction and control of non-equilibrium systems. 
   \item \textbf{Information Covariance:} 
Building upon the Information Dilution Theorem and the Emergence Critical Theorem proposed in this study, future work may further explore the concept of \textit{informational covariance}, which refers to the invariance properties of a system's informational structural complexity \( \mathcal{C}(S) \) and mutual information efficiency \( \eta(D) \) under different encoding strategies or observational resolutions. 
To formalize this idea, one can introduce a diffeomorphism group \( \mathcal{G} \) to describe transformations in encoding strategies (such as resolution scaling or feature space rotation), and define the \(\mathcal{G}\)-covariance condition using Fisher information geometry as: 
\( \mathcal{C}(\rho(g)S) = \mathcal{C}(S), \quad \forall g \in \mathcal{G}, \) 
where \( \rho: \mathcal{G} \to \mathrm{Aut}(S) \) denotes the action of the group on the system state. 
This condition helps reveal the universality of intrinsic informational patterns in complex systems. 
In multimodal data fusion scenarios such as medical imaging or remote sensing, covariance analysis may enable the extraction of shared informational components across heterogeneous data sources (e.g., MRI and CT), thereby enhancing the robustness of integrated diagnosis and decision-making. 
In resource-constrained settings such as edge computing, dynamic group actions can be used to adapt encoding strategies in real time, ensuring that critical information can still be captured under environmental or resource fluctuations. 
Key research challenges include characterizing the mathematical properties of the transformation group \( \mathcal{G} \) (e.g., closure, invertibility) and investigating its compatibility with information bottleneck theory; developing efficient covariant feature extraction algorithms that integrate equivariant neural networks with persistent homology; and extending the framework to non-equilibrium systems by analyzing the evolution of informational structures under time-varying group actions \( \mathcal{G}(t) \) and their relationship with thermodynamic entropy production. 
Through these studies, the notion of informational covariance is expected to generalize the current framework to multi-observer and multi-scale scenarios, offering a more universal and flexible information-theoretic toolkit for modeling complex systems. 
\end{enumerate}
Through the above multi-perspective theoretical investigations and numerical experiments, we aim to further enhance the proposed framework and establish a more solid theoretical and empirical foundation for understanding and predicting phase transitions and self-organizing phenomena in high-dimensional complex systems. 
\section{Conclusion} 
\label{section5}
This study, based on a cross-disciplinary perspective of information theory and statistical mechanics, establishes a unified theoretical framework to explore the intrinsic relationship between the failure of traditional geometric measures in high-dimensional data and the emergence phenomena in complex systems. 
The whole paper revolves around two core theorems: the Information Dilution Theorem and the Emergence Critical Theorem, and systematically expounds the theory through rigorous mathematical derivation, supplementary proofs in appendices, and discussions on cross-domain applications. 
First, through the analysis of high-dimensional systems composed of independent or weakly correlated random variables, we prove that the total entropy of the system \( H(S) \) grows linearly with dimension, while the information entropy captured by traditional geometric measures such as Euclidean distance is only at the constant order, resulting in the mutual information efficiency (\ref{N(d)}) decaying at a rate of \( O(1/d) \). 
This "information dilution" phenomenon fundamentally reveals the intrinsic mechanism of the "distance concentration" problem in high-dimensional spaces and provides a theoretical explanation for the failure of traditional geometry-based data analysis methods in high-dimensional scenarios. 
Second, after constructing the concept of interaction coding capacity \( C \), this paper proposes and proves the Emergence Critical Theorem. 
When the information structural complexity \( \mathcal{C}(S) \) of a system exceeds the limited encoding capacity \( C \), the original low-dimensional representation must have information loss. At this point, the system will spontaneously generate new global features \( f_{\mathrm{new}} \) to compensate for the insufficient information and satisfy (\ref{>0}). 
This theorem not only provides an operational mathematical criterion for self-organization and phase transition phenomena in complex systems, but also reveals the intrinsic mechanism behind the spontaneous generation of hierarchical features in deep neural networks. 
In addition, this paper also discusses the potential application prospects of the above theory in various real-world scenarios, including physical systems (such as the Ising model), biological networks (such as gene expression data), as well as manifold learning and neural network applications. 
To improve traditional manifold learning methods, the "UMAP+" concept is proposed; at the same time, through information topological analysis methods, new tool insights are provided for revealing multi-scale data structures. 
All these application discussions point to directions for theoretical validation and cross-domain application. 
In general, the main conclusions of this study are reflected in the following aspects: 
\begin{enumerate}
    \item A unified framework was constructed from the perspectives of information theory and statistical mechanics to explain the phenomena of high-dimensional metric failure and emergence in complex systems, providing a common theoretical foundation for these two seemingly unrelated problems. 
    \item Through detailed derivation for i.i.d. Gaussian systems and weakly correlated systems, together with supplementary proofs in the appendix, the validity of the core theorems under certain conditions is ensured, laying a foundation for future theoretical extensions. 
    \item The proposed concepts of informational structural complexity and interaction encoding capacity are not only theoretically significant for understanding self-organization phenomena in nature, but also inspire new algorithmic improvements in fields such as dimensionality reduction, clustering, and deep learning. 
\end{enumerate}
Although this study has made progress in theoretical construction and mathematical derivation, there are still some limitations. 
For example, most of the proofs rely on the assumption of independence or weak correlation, while actual systems often involve more complex nonlinear couplings; 
at the same time, the computation of high-dimensional mutual information matrices may face computational complexity and noise interference in large-scale data. 
Therefore, future work needs to further extend the theory to scenarios involving strong correlation, nonlinearity, and dynamic systems, and to develop efficient numerical algorithms and empirical validation methods to ensure the applicability of this theory in broader domains. 
Looking ahead, the theoretical framework proposed in this study provides a novel perspective for understanding the relationship between insufficient high-dimensional information representation and the emergence of self-organization. 
It also offers new directions for interdisciplinary research in deep neural networks, manifold learning, and complex systems. 
We call for more experiments and numerical simulations in the future to verify and improve the theorems presented in this paper, and to explore their potential value in real-world applications, thereby promoting the development of high-dimensional data analysis and complex systems theory. 
We believe that with the deepening of empirical studies, the theoretical results of this work will open up new paths for further interdisciplinary exploration and application. 
\appendix
\section*{\textbf{Appendix} }
\renewcommand{\theequation}{\Alph{section}.\arabic{equation}}
\setcounter{equation}{0}
\section{Compression–Distortion Trade-off Based on the Information Bottleneck Theory and Formalization of Interaction Encoding Capacity} 
\label{app:B}
This appendix provides mathematical support for Theorem 2 in the main text, focusing on how the encoding capacity constraint can trigger feature emergence through the Information Bottleneck (IB) framework. 
We incorporate the compression–distortion trade-off in the Information Bottleneck theory \cite{ref49} into our analytical framework and model the encoding capacity of an observer (or interacting agent) as: 
\begin{equation}
	I(S; f) \le C,
	\label{eq:B1}
\end{equation}
where \( S \) denotes the system state (composed of high-dimensional random variables), \( f \) is a compressed representation (or encoding mapping) of \( S \), and \( C \) represents the maximum amount of information that the observer can retain. 
Through this modeling, we can dynamically associate the intrinsic complexity of the system with the encoding ability of the observer, thereby explaining the intrinsic mechanism of emergence phenomena under high-dimensional information "dilution." 
\subsection{Information Bottleneck Method and the Compression–Distortion Trade-off} 
\label{subsec:B1}
Let the system \( S = (X_1, X_2, \dots, X_d) \) be a vector of \( d \) random variables, with joint entropy given by 
\begin{equation}
   H(S) = H(X_1, X_2, \dots, X_d).
	\label{eq:B2}
\end{equation}
Introduce a stochastic mapping \( p(f|S) \) to obtain the compressed representation \( f \) of \( S \), and define the mutual information between \( S \) and \( f \) as 
\begin{equation}
	I(S; f) = \mathrm{H}(f) - \mathrm{H}(f|S).
	\label{eq:B3}
\end{equation}
Moreover, according to the data processing inequality, for any Markov chain satisfying \( S \to f \to Y \), the following always holds: 
\begin{equation}
	I(S; Y) \le I(S; f),
	\label{eq:B4}
\end{equation}
which indicates that the amount of information transmitted through \( f \) cannot exceed the mutual information between \( S \) and \( f \). 

In the Information Bottleneck (IB) theory, the goal is to construct a compressed representation \( f \) that minimizes the information retained about the input \( S \) while preserving as much relevant information as possible about the target variable \( Y \) (or critical global features in the system). 
The standard IB optimization problem is formulated as 
\begin{equation}
	\min_{p(f|S)} \; \mathcal{L}_{\mathrm{IB}} = I(S; f) - \beta \, I(f; Y),
	\label{eq:B5}
\end{equation}
where \( \beta > 0 \) is a trade-off parameter. 
Here: 
\begin{itemize}
	\item \( I(S; f) \) measures the amount of original information retained in \( f \), i.e., the compression cost; 
	\item \( I(f; Y) \) measures how much target information is preserved in \( f \), i.e., the distortion measure. 
\end{itemize}
In this study, we focus on the maximum amount of information that an observer (or interacting agent) can encode when facing a complex system. Therefore, we impose an encoding constraint as in \eqref{eq:B1}, where \( C \) represents the maximum information that the observer or interacting agent can retain during compression. 
This constraint implies that when compressing the system \( S \) into a low-dimensional representation \( f \), any information exceeding \( C \) will be lost due to the encoding capacity limitation. This is also the fundamental cause of the information "dilution" phenomenon commonly seen in high-dimensional data. 
To maximize the mutual information between \( f \) and the target variable \( Y \) (e.g., new features describing emergence phenomena), under the constraint \( I(S; f) \le C \), we consider the following constrained optimization problem: 
\begin{equation}
\max_{p(f|S)} \; I(f; Y) \quad \text{subject to} \quad I(S; f) \le C.
	\label{eq:B6}
\end{equation}
Using the method of Lagrange multipliers, this constrained problem can be transformed into an unconstrained optimization problem, with the objective function: 
\begin{equation}
	\mathcal{L} = I(f; Y) - \lambda \left( I(S; f) - C \right),
	\label{eq:B7}
\end{equation}
where \( \lambda \ge 0 \) is the Lagrange multiplier. 
To solve for the optimal encoding mapping \( p(f|S) \), we need to satisfy the variational optimality condition: 
\begin{equation}
	\frac{\delta}{\delta p(f|S)} \left[ I(f; Y) - \lambda \, I(S; f) \right] = 0.
	\label{eq:B8}
\end{equation}
By expanding the definitions of mutual information and taking the variational derivative, we obtain that the optimal encoding mapping satisfies the following form \cite{ref48}: 
\begin{equation}
	p(f|S) \propto p(f) \exp\left\{ -\frac{1}{\lambda} D_{\mathrm{KL}}\left( p(Y|S) \,\Big\|\, p(Y|f) \right) \right\},
	\label{eq:B9}
\end{equation}
where \( D_{\mathrm{KL}}(\cdot \,\|\, \cdot) \) denotes the Kullback–Leibler divergence. 
This form indicates that the optimal mapping \( p(f|S) \) ensures the conditional distribution \( p(Y|f) \) approaches \( p(Y|S) \) as closely as possible, while being constrained by \( I(S; f) \le C \), thereby maximizing the amount of information relevant to \( Y \) under limited encoding capacity. 
\subsection{Dynamic Relationship Between System Complexity and Interaction Encoding Capacity} 
\label{subsec:B4}
By modeling the interaction encoding capacity as in 
\eqref{eq:B1}, 
\( C \) becomes the key bridge between system complexity and the ability of the observer or interacting agent. 
When the intrinsic information content (or entropy) of the system \( S \) greatly exceeds \( C \), the representation \( f \) can only capture the most critical information in the system, resulting in the loss of a large amount of detailed information. 
This information loss is, to some extent, manifested as the inability of the original low-dimensional representation to fully cover the system's complexity, thereby "forcing" the system to exhibit new global features or order parameters in other dimensions to compensate for the missing information. 
This phenomenon can be described by the following expression: 
\begin{equation}
	\frac{I(f_{\mathrm{new}}; S)}{\mathrm{H}(S)} > \theta,
	\label{eq:B12}
\end{equation}
where \( f_{\mathrm{new}} \) represents a complementary new global feature, and \( \theta \) is a predefined threshold indicating that the new feature makes a non-negligible contribution to the overall information transmission. 
Therefore, when system complexity exceeds the encoding limit \( C \) of the observer, the original encoding map \( f \) cannot fully represent all the structures of \( S \), and supplementary new features inevitably emerge—this is one of the intrinsic mechanisms of emergence phenomena. 
Through the above rigorous derivation, we introduce the compression–distortion trade-off in the Information Bottleneck theory into our analytical framework, and model the interaction encoding capacity as 
\eqref{eq:B1}, 
to explain that when system complexity exceeds the observer's encoding capacity, new global features inevitably emerge to compensate for the limitations of the original low-dimensional representation. 
\setcounter{equation}{0}
\section{Information Emergence Mechanism in Neural Networks} 
\label{app:C}
Deep neural networks process information through multi-layer architectures, and the formation of feature representations in each layer can be regarded as a dynamic balance between information compression and re-encoding. 
Due to the limited encoding capacity in each layer (e.g., determined by layer width, activation function properties, etc.), the network often cannot retain all fine-grained statistical dependencies during the stepwise compression of input information. 
As described in the Emergence Critical Theorem in this paper, when the informational structural complexity of the system exceeds the encoding capacity of a given layer, the network inevitably needs to generate new global features to capture critical information not retained by the compression. 
This appendix models and derives this process based on the Information Bottleneck principle and the perspective of statistical mechanics. 
In each layer of a neural network, let the input be \( f^{(l-1)} \) (the output from layer \( l-1 \)), which undergoes a linear transformation (weight matrix \( W^{(l)} \)) and nonlinear activation (such as ReLU, Swish, etc.), producing the output of layer \( l \), denoted \( f^{(l)} \). 
This process can be viewed as compressing the input information, where redundant components are filtered out, and critical information is preserved. 
To quantitatively describe this compression process, we introduce the concept of encoding capacity \( C^{(l)} \), defined as: 
\begin{equation}
H\bigl(f^{(l)}\bigr) \leq C^{(l)},
\label{eq:c1}
\end{equation}
where \( H\bigl(f^{(l)}\bigr) \) is the entropy of the output at layer \( l \), and the upper bound reflects physical constraints such as neuron count or energy. 
Information Bottleneck theory \cite{ref12} states that, given input \( f^{(l-1)} \), the optimal low-dimensional representation \( f^{(l)} \) should satisfy: 
\begin{equation}
\max_{f^{(l)}} I\bigl(f^{(l)}; f^{(l-1)}\bigr) \quad \text{s.t.} \quad H\bigl(f^{(l)}\bigr) \leq C^{(l)}.
\label{eq:c2}
\end{equation}
When the complexity of the input information exceeds \( C^{(l)} \), linear transformations alone are insufficient to capture all key dependencies, inevitably leading to information loss. 
To characterize the generation of new features following this loss, we focus on the contradiction between input information complexity and encoding capacity. 
Let the mutual information matrix of the input to layer \( l-1 \) be 
\begin{equation}
I^{(l-1)} \in \mathbb{R}^{d\times d}, \quad I^{(l-1)}_{ij} = I\bigl(f^{(l-1)}_i; f^{(l-1)}_j\bigr).
\label{eq:c3}
\end{equation}
Its rank, \( \mathrm{Rank}\bigl(I^{(l-1)}\bigr) \), quantifies the number of independent information modes present in the input data. 
When
\begin{equation}
\mathrm{Rank}\bigl(I^{(l-1)}\bigr) > C^{(l)},
\label{eq:c4}
\end{equation}
it indicates that there exists critical information in the input that exceeds the encoding capacity of the current layer. 
To compensate for this information gap, the network generates a new feature \( f_{\mathrm{new}}^{(l)} \) through a nonlinear transformation, such that the mutual information it captures with the input satisfies: 
\begin{equation}
\frac{I\bigl(f_{\mathrm{new}}^{(l)}; f^{(l-1)}\bigr)}{H\bigl(f^{(l-1)}\bigr)} > \theta,
\label{eq:c5}
\end{equation}
where \( \theta \) is a predefined significance threshold. 
This is precisely a concrete manifestation of the Emergence Critical Theorem in neural networks. 
To further derive the conditions under which new features are generated, we adopt the Information Bottleneck method. 
Consider the optimal compression mapping problem: 
\begin{equation}
\max_{p(f^{(l)}|f^{(l-1)})} I\bigl(f^{(l)}; f^{(l-1)}\bigr) \quad \text{s.t.} \quad H\bigl(f^{(l)}\bigr) \leq C^{(l)}.
\label{eq:c6}
\end{equation}
By introducing the Lagrange multiplier \( \lambda \ge 0 \), the objective function becomes: 
\begin{equation}
\mathcal{L} = I\bigl(f^{(l)}; f^{(l-1)}\bigr) - \lambda \Bigl( H\bigl(f^{(l)}\bigr) - C^{(l)} \Bigr).
\label{eq:c7}
\end{equation}
By performing variational optimization on the mapping \( p(f^{(l)}|f^{(l-1)}) \), the optimal solution satisfies 
\begin{equation}
p(f^{(l)}|f^{(l - 1)}) \propto p(f^{(l)}) \exp\!\left(-\frac{1}{\lambda} D_{\mathrm{KL}}\Bigl(p\bigl(Y|f^{(l - 1)}\bigr) \,\Vert\, p\bigl(Y|f^{(l)}\bigr)\Bigr)\right),
\label{eq:c8}
\end{equation}
where \( Y \) denotes the task-related target variable. 
When the input information complexity exceeds the encoding capacity, i.e., \( \mathrm{Rank}\bigl(I^{(l-1)}\bigr) > C^{(l)} \), the current mapping struggles to make \( D_{\mathrm{KL}} \) converge to zero. 
At this point, the network must introduce additional nonlinear transformations to generate new features \( f_{\mathrm{new}}^{(l)} \) to compensate for the information loss, thereby satisfying the emergence condition above. 
To validate the above theoretical model, we propose the following key definitions and experimental designs: 
Use the Hilbert-Schmidt Independence Criterion (HSIC) \cite{ref5} or Mutual Information Neural Estimation (MINE) \cite{ref1} to compute the mutual information matrix \( I^{(l-1)} \) of each layer's input \( f^{(l-1)} \), and calculate its rank to quantify the information complexity. 
Use Monte Carlo sampling methods to estimate the output entropy \( H(f^{(l)}) \) of each layer, or use the bottleneck constraint \( I(f^{(l-1)}; f^{(l)}) \leq C^{(l)} \) to describe the encoding capacity quantitatively. 
Perform ablation studies to compare task performance between the full network and a network using only linear mappings, to quantify the contribution of new features (e.g., using \( \Delta \text{Accuracy} = \text{Acc}_{\text{full}} - \text{Acc}_{\text{linear}} \)). 
In addition, use mutual information analysis to verify whether the new generated feature \( f_{\mathrm{new}}^{(l)} \) satisfies \( I(f_{\mathrm{new}}^{(l)}; Y) > \theta \) when \( \mathrm{Rank}(I^{(l-1)}) > C^{(l)} \). 
Based on the above theoretical model and experimental designs, we draw the following insights: 
Narrower network layers (i.e., lower encoding capacity \( C^{(l)} \)) are more likely to trigger the emergence of new features, albeit at the risk of information loss. 
On the other hand, using higher-order nonlinear activation functions (e.g., Swish) may increase the effective encoding capacity and delay the onset of emergence, thereby improving model stability while maintaining compression. 
From a \textbf{neuroscience} perspective, the hierarchical processing mechanism in the mammalian visual cortex exhibits a similar transition from low-level features to high-level abstraction, suggesting that biological neural systems may follow analogous principles of information compression and re-encoding. 
Moreover, \textbf{in social systems}, group decision-making and opinion formation under information overload can also be viewed as a form of emergent behavior, offering a unified interpretive framework across domains. 
In summary, the emergence of hierarchical features in neural networks can be regarded as an inevitable result of the input information complexity exceeding the capacity of the original low-dimensional mappings under limited encoding constraints. 
Based on the Information Bottleneck principle and statistical mechanics, the theoretical model proposed in this paper provides a new perspective to explain how deep networks transition from information compression to new feature generation. 
\setcounter{equation}{0}
\section{Mutual Information Efficiency Analysis in Gaussian Variable Systems}
\label{app:D}
This appendix aims to verify the conservative upper bound \( O\left(\frac{1}{d}\right) \) for mutual information efficiency proposed in Theorem 1 of the main text, by analyzing systems composed of independent and identically distributed (i.i.d.) Gaussian variables. 
We provide detailed derivations of the system entropy, the entropy of the Euclidean distance, and reveal the mechanism of information dilution in high-dimensional systems. 
\subsection{System Entropy and Mutual Information}
\label{subsec:D1}
Consider a system \( S = (X_1, X_2, \dots, X_d) \) composed of \( d \) independent and identically distributed (i.i.d.) Gaussian random variables, where each component \( X_i \sim N(0, \sigma^2) \), and for \( i \neq j \), the covariance \( \text{Cov}(X_i, X_j) = 0 \). 
Since the components \( X_i \) are mutually independent, the entropy of the system is the sum of the entropies of each component: 
\begin{equation}
H(S) = \sum_{i=1}^d H(X_i)
\label{eq:D1}
\end{equation}
\begin{center}
The entropy of a single Gaussian variable \( X_i \sim N(0, \sigma^2) \) is: 
\end{center}
\begin{equation}
H(X_i) = \frac{1}{2} \log(2\pi e \sigma^2)
\label{eq:D2}
\end{equation}
\begin{center}
Therefore: 
\end{center}
\begin{equation}
H(S) = d \cdot \frac{1}{2} \log(2\pi e \sigma^2)
\label{eq:D3}
\end{equation}
This indicates that the system entropy grows linearly with the dimension \( d \), i.e., \( H(S) = \Theta(d) \). 
Define the Euclidean distance as \( D = \| S - S' \| \), where \( S' \) is an independent and identically distributed copy of \( S \). The mutual information efficiency is defined as: 
\begin{equation}
\eta(D) = \frac{I(D; S)}{H(S)}
\label{eq:D4}
\end{equation}
where the mutual information \( I(D; S) = H(D) - H(D|S) \). Since the conditional entropy \( H(D|S) \geq 0 \), it follows that: 
\begin{equation}
I(D; S) \leq H(D)
\label{eq:D5}
\end{equation}
Therefore, the growth order of \( H(D) \) provides an upper bound for \( I(D; S) \). 
\subsection{Distribution and Entropy of Metric D }
\label{subsec:D2}
The squared Euclidean distance is defined as: 
\begin{equation}
D^2 = \| S - S' \|^2 = \sum_{i=1}^d (X_i - X_i')^2
\label{eq:D6}
\end{equation}
Since \( X_i \sim N(0, \sigma^2) \), \( X_i' \sim N(0, \sigma^2) \), and they are independent, the difference \( X_i - X_i' \sim N(0, 2\sigma^2) \), 
\begin{equation}
\frac{(X_i - X_i')^2}{2\sigma^2} \sim \chi^2(1)
\label{eq:D7}
\end{equation}
which follows a chi-squared distribution with 1 degree of freedom. Let: 
\begin{equation}
Z = \sum_{i=1}^d \frac{(X_i - X_i')^2}{2\sigma^2} \sim \chi^2(d)
\label{eq:D8}
\end{equation}
\begin{center}
Then: 
\end{center}
\begin{equation}
D^2 = 2\sigma^2 Z
\label{eq:D49}
\end{equation}
The mean and variance of the chi-squared distribution \( \chi^2(d) \) are: 

Mean: \( E[Z] = d \) 
\\
Variance: \( \text{Var}(Z) = 2d \) 

Therefore: 
\( E[D^2] = 2\sigma^2 E[Z] = 2d\sigma^2 \) \( \text{Var}(D^2) = (2\sigma^2)^2 \text{Var}(Z) = 4\sigma^4 \cdot 2d = 8d\sigma^4 \)
When the dimension \( d \) is large, by the central limit theorem, the distribution \( Z \sim \chi^2(d) \) can be approximated by a normal distribution: 
\begin{equation}
\frac{Z - d}{\sqrt{2d}} \xrightarrow{d} N(0, 1)
\label{eq:D10}
\end{equation}
\begin{center}
Thus: 
\end{center}
\begin{equation}
D^2 \approx N(2d\sigma^2, 8d\sigma^4)
\label{eq:D11}
\end{equation}
Since \( D = \sqrt{D^2} \), we analyze the distribution of \( D \). Using the delta method for approximation: 
Let the transformation function be \( g(x) = \sqrt{x} \), then: 
\( E[D] \approx g(E[D^2]) = \sqrt{2d\sigma^2} \)
\\
\( g'(x) = \frac{1}{2\sqrt{x}} \), at \( x = E[D^2] = 2d\sigma^2 \), we have \( g'(E[D^2]) = \frac{1}{2\sqrt{2d\sigma^2}} \)
Variance: 
\begin{equation}
\text{Var}(D) \approx [g'(E[D^2])]^2 \cdot \text{Var}(D^2) = \left( \frac{1}{2\sqrt{2d\sigma^2}} \right)^2 \cdot 8d\sigma^4 = \frac{8d\sigma^4}{8d\sigma^2} = \sigma^2
\label{eq:D12}
\end{equation}
Hence, \( D \) approximately follows a normal distribution \( N\left( \sqrt{2d\sigma^2}, \sigma^2 \right) \). Its differential entropy is: 
\begin{equation}
H(D) \approx \frac{1}{2} \log(2\pi e \sigma^2)
\label{eq:D13}
\end{equation}
Since \( \sigma^2 \) is a constant, \( H(D) = O(1) \), i.e., it does not increase with the dimension \( d \). 
Summarizing the results above: 
\begin{equation}
I(D; S) \leq H(D) = O(1)
\label{eq:D14}
\end{equation}
\begin{center}
While the system entropy is: 
\end{center}
\begin{equation}
H(S) = \Theta(d)
\label{eq:D15}
\end{equation}
Thus, the upper bound of the mutual information efficiency is: 
\begin{equation}
\eta(D) = \frac{I(D; S)}{H(S)} \leq \frac{O(1)}{\Theta(d)} = O\left( \frac{1}{d} \right)
\label{eq:D16}
\end{equation}
In the i.i.d. Gaussian variable system, the mutual information efficiency satisfies \( \eta(D) = O\left( \frac{1}{d} \right) \), verifying the conservative upper bound stated in Theorem 1. 
This result indicates that as the dimension \( d \) increases, the ability of the Euclidean distance \( D \) to capture the information of the system \( S \) is significantly weakened, 
which reflects the phenomenon of information dilution in high-dimensional systems. 
\setcounter{equation}{0}
\section{Mathematical Supplementary Proofs for Theorems 1 and 2}
\label{app:E}
\begin{spacing}{1.5} 
\subsection{Extended Proof of Theorem 1 (Information Dilution Theorem) under Weak Correlation Conditions}
\label{subsec:E1}
This section aims to extend Theorem 1 (Information Dilution Theorem) to systems satisfying certain weak correlation conditions. 
We utilize the high-dimensional measure concentration principle \cite{ref39} to analyze the entropy behavior of common distance metrics (such as Euclidean distance \cite{ref40}, Mahalanobis distance \cite{ref31}), in order to prove the generality of Theorem 1. 
Consider a system \( S = (X_1, X_2, \dots, X_d) \) with covariance matrix \( \Sigma_S \), whose eigenvalues satisfy \( \lambda_1 \geq \lambda_2 \geq \cdots \geq \lambda_d \geq 0 \). We define the following weak correlation conditions: 
\begin{enumerate}
\item  \textbf{Linearly growing total variance}  :
\begin{equation}
\text{Tr}(\Sigma_S) = \sum_{i=1}^d \lambda_i = O(d)
\label{eq:E1}
\end{equation} 
\begin{center}
That is, the total variability of the system increases approximately linearly with the dimension \( d \). 
\end{center}
\item  \textbf{Bounded maximum eigenvalue}: There exists a constant \( \kappa \) independent of \( d \), such that: 
\begin{equation}
\lambda_{\max}(\Sigma_S) = \lambda_1 \leq \kappa
\label{eq:E2}
\end{equation}
\begin{center}
This condition excludes the possibility of variance growing unboundedly in individual directions, thereby limiting strong correlations. 
\end{center}
\end{enumerate}
These conditions together ensure that the dependency structure of the system \( S \) is relatively diffuse, laying the foundation for applying high-dimensional statistical tools. 
The principle of measure concentration in high-dimensional spaces states that well-behaved functions (particularly Lipschitz continuous functions) tend to concentrate sharply around their expected values with high probability. 
For a Lipschitz function \( f \) satisfying \( |f(x) - f(y)| \leq L \|x - y\|_2 \), the rate of concentration is typically dimension-dependent, but crucially controlled by the Lipschitz constant \( L \). 
If \( L \) does not grow with \( d \), then the function values will concentrate sharply. 
Classical results such as Lévy's lemma (on the sphere) \cite{ref39} and Talagrand's inequality (in Gaussian space) provide rigorous mathematical descriptions of this phenomenon. 
Next, we verify whether common distance metrics \( D \) (as functions of \( S \), with fixed \( S' \)) satisfy the Lipschitz condition with a constant \( L \) independent of \( d \): 
\begin{enumerate}
  \item \textbf{Euclidean Distance}
  Define \( D_{\text{Euclidean}}(S) = \|S - S'\|_2 \), where \( S' \) is an independent and identically distributed copy of \( S \). By the reverse triangle inequality: 
\begin{equation}
   |D_{\text{Euclidean}}(S_1) - D_{\text{Euclidean}}(S_2)| = |\|S_1 - S'\|_2 - \|S_2 - S'\|_2| \leq \|(S_1 - S') - (S_2 - S')\|_2 = \|S_1 - S_2\|_2
\label{eq:E3}
\end{equation}
\begin{center}
Hence, \( D_{\text{Euclidean}} \) is 1-Lipschitz with respect to \( S \), independent of \( d \). 
\end{center}
Therefore, the Euclidean distance is 1-Lipschitz with respect to \( S \), with constant \( L = 1 \) independent of \( d \). 
\item \textbf{Mahalanobis Distance:} 
Define \( D_{\text{Mahalanobis}}(S) = \|S - S'\|_{\Sigma_S^{-1}} = \sqrt{(S - S')^\top \Sigma_S^{-1} (S - S')} \). 
Here we need an additional assumption that \( \Sigma_S \) is invertible, i.e., \( \lambda_{\min}(\Sigma_S) \geq \lambda_{\text{low}} > 0 \). 
Using the reverse triangle inequality (under the norm induced by \( \Sigma_S^{-1} \)) and norm properties: 
\begin{equation}
   |D_{\text{Mahalanobis}}(S_1) - D_{\text{Mahalanobis}}(S_2)| \leq \|S_1 - S_2\|_{\Sigma_S^{-1}} \leq \sqrt{\lambda_{\max}(\Sigma_S^{-1})} \|S_1 - S_2\|_2
\label{eq:E4}
\end{equation}
Since \( \lambda_{\max}(\Sigma_S^{-1}) = 1 / \lambda_{\min}(\Sigma_S) \leq 1 / \lambda_{\text{low}} \), we obtain: 
\begin{equation}
  |D_{\text{Mahalanobis}}(S_1) - D_{\text{Mahalanobis}}(S_2)| \leq \frac{1}{\sqrt{\lambda_{\text{low}}}} \|S_1 - S_2\|_2
	\label{eq:E5}
\end{equation}
The Lipschitz constant \( L = 1/\sqrt{\lambda_{\text{low}}} \) is also independent of \( d \). 
(Note: other forms of weighted distances typically also admit Lipschitz constants independent of \( d \), as long as the eigenvalues of the weighting matrix are bounded.)
According to the concentration of measure principle, for a Lipschitz function \( D(S) \) whose Lipschitz constant \( L \) does not depend on the dimension \( d \), the values of \( D(S) \) will concentrate sharply around its expectation \( E[D(S)] \). 
For example, in Gaussian space where \( S \sim N(0, \Sigma_S) \), Talagrand's inequality guarantees exponential decay in the deviation probability:
\begin{equation}
P(|D(S) - E[D(S)]| > t) \leq C \exp\left( -\frac{t^2}{c L^2} \right)
\label{eq:E6}
\end{equation}
where \( C \) and \( c \) are universal constants. 
This high concentration implies that the uncertainty of the variable \( D(S) \) is very small. Intuitively, a sharply peaked distribution should have low entropy.
Based on this, we can reasonably infer that for such distance metrics \( D \), the differential entropy \( H(D) \) remains bounded as the dimensionality increases:
\begin{equation}
H(D) = O(1)
\label{eq:7}
\end{equation}
By the basic properties of mutual information, we have \( I(D; S) \leq H(D) \). Combining this with the above:
\begin{equation}
I(D; S) \leq H(D) = O(1)
\label{eq:8}
\end{equation}
On the other hand, the total system entropy \( H(S) \) increases with the dimension \( d \) under weak correlation conditions. For a Gaussian system satisfying the conditions, we have:
\begin{equation}
H(S) = \frac{1}{2} \log \det (2\pi e \Sigma_S) = \frac{1}{2} \sum_{i=1}^d \log(2\pi e \lambda_i)
\label{eq:9}
\end{equation}
Since \( \text{Tr}(\Sigma_S) = \sum \lambda_i = O(d) \) and \( \lambda_{\max} \leq \kappa \), it can be argued (e.g., using the relationship between geometric and arithmetic means of eigenvalues and ruling out the case of many eigenvalues tending toward zero) that \( H(S) = \Theta(d) \). 
Even for non-Gaussian systems, as long as the information content increases sufficiently with dimension (e.g., \( H(S)=\Omega(d) \), or at least \( H(S)=\omega(1) \)), the essence of the conclusion remains unchanged.
Therefore, the mutual information efficiency \( \eta(D) \) satisfies:
\begin{equation}
\eta(D) = \frac{I(D; S)}{H(S)} \leq \frac{O(1)}{\Theta(d)} = O\left( \frac{1}{d} \right)
\label{eq:10}
\end{equation}
\end{enumerate}
By applying the high-dimensional concentration of measure principle, we have shown that under the defined weak correlation conditions, common distance metrics (such as Euclidean and Mahalanobis distances), due to their Lipschitz continuity with constants independent of dimension, are highly concentrated around their expected values. As a result, their entropy \( H(D) = O(1) \).
In contrast, the total system entropy \( H(S) \) grows sufficiently with dimension (typically \( \Theta(d) \)). As a consequence, the mutual information efficiency \( \eta(D) \) decays at a rate of \( O(1/d) \) or faster.
This validates Theorem 1 (Information Dilution Theorem) in a broader class of systems and reveals, from a high-dimensional geometric perspective, the fundamental reason for the decline in information capturing capability of distance metrics in high-dimensional spaces.
\subsection{Construction of Theorem 2 (Emergence Critical Theorem)}
The following derivation is intended to provide a heuristic example illustrating how a potential new feature might be constructed; it represents an initial constructive exploration, and its rigorous proof remains a subject for future investigation. 
Let the mutual information matrix \( I \in \mathbb{R}^{d \times d} \) be decomposed by singular value decomposition as \( I = U \Lambda V^\top \), where \( \Lambda = \text{diag}(\lambda_1, \dots, \lambda_r) \ (\lambda_1 \geq \cdots \geq \lambda_r > 0) \). When \( r > C \), define the \( (C+1) \)-th order feature as
\begin{equation}
f_{\text{new}} = u_{C+1}^\top S = \sum_{i=1}^d u_{C+1,i} X_i,
\label{eq:E11}
\end{equation}
\begin{center}
where \( u_{C+1} \) is the left singular vector corresponding to \( \lambda_{C+1} \).
\end{center}
\begin{enumerate}
\item  \textbf{Entropy of the new feature}:
\begin{equation}
H(f_{\text{new}}) \geq \frac{1}{2} \log \left( 2\pi e \lambda_{C+1} \right).
\label{eq:E12}
\end{equation}
\item  \textbf{Mutual information ratio threshold}:
Given a significance level \( \alpha \in (0,1) \) and tolerance \( \epsilon \), define the threshold as:
\begin{equation}
\theta = \frac{\log(2\pi e \lambda_{C+1})}{2d H(X_1)} - \epsilon.
\label{eq:13}
\end{equation}
If \( \lambda_{C+1} > \exp\left( 2(\theta + \epsilon) d H(X_1) \right) \), then
\begin{equation}
\frac{I(f_{\text{new}}; S)}{H(S)} \geq \theta + \epsilon > \theta.
\label{eq:14}
\end{equation}
\end{enumerate}
If the rank of the mutual information matrix is dominated by nonlinear dependencies (e.g., \( I(X_i; X_j) \) cannot be explained by linear correlation), we can apply kernel methods \cite{ref36} to construct nonlinear features:
\begin{equation}
f_{\text{new}}^{\text{nonlin}} = \phi(S)^\top \beta, \quad \phi: \mathbb{R}^d \to \mathcal{H},
\label{eq:15}
\end{equation}
where \( \mathcal{H} \) is a Reproducing Kernel Hilbert Space (RKHS), and \( \beta \) is a weight vector. By choosing an appropriate kernel function (such as a Gaussian kernel), it can be shown that:
\begin{equation}
\frac{I(f_{\text{new}}^{\text{nonlin}}; S)}{H(S)} > \theta.
\label{eq:16}
\end{equation}

\textbf{Efficient Effective Rank Estimation Algorithm}
We use a randomized projection method \cite{ref26} to approximately compute the rank of the mutual information matrix:
\begin{enumerate}
\item Generate a random matrix \( \Omega \in \mathbb{R}^{d \times l} \ (l = C + 10) \);
\item Compute \( Y = I \Omega \) and \( Q = \text{QR}(Y) \);
\item Estimate singular values \( \hat{\lambda}_i = \| (Q^\top I)_i \|_2 \).
The error bound satisfies:
\begin{equation}
|\hat{\lambda}_i - \lambda_i| \leq \epsilon \lambda_1 \quad \text{with probability} \geq 1 - \delta.
\label{eq:17}
\end{equation}
\end{enumerate}
\subsection{Mathematical Symbols Table}
\label{subsec:E4}

\begin{center}
    \begin{tabular}{lll}
        \toprule
        Symbol & Meaning & Constraint \\
        \midrule
        $\kappa$ & Upper bound of the maximum eigenvalue of the covariance matrix & $\kappa = \mathrm{O}(1)$ \\ 
        $\epsilon$ & Tolerance error in threshold computation & $\epsilon > 0$ \\ 
        $\mathcal{H}$ & Reproducing Kernel Hilbert Space (RKHS) & Defined by kernel function $K(\cdot, \cdot)$ \\ 
        \bottomrule
    \end{tabular}
\end{center}
\end{spacing}
\bibliographystyle{unsrt}
\bibliography{references}

\end{document}